\documentclass[showpacs,prl,floatfix,twocolumn,amsmath,a4]{revtex4}

\usepackage{graphicx}

\begin{document} 

\title{Multiferroicity in rare-earth nickelates RNiO$_3$}

\author{Gianluca Giovannetti$^{1,2,3}$, Sanjeev Kumar$^{1,2}$, Daniel Khomskii$^4$, Silvia Picozzi$^{3}$ and Jeroen van den Brink$^{1,5,6}$}

\address{
$^1$Institute Lorentz for Theoretical Physics, Leiden University, 2300 RA Leiden, The Netherlands\\ 
$^2$Faculty of Science and Technology and MESA+ Research Institute, University of Twente, The Netherlands\\ 
$^3$Consiglio Nazionale delle Ricerche - Istituto Nazionale per la Fisica della Materia (CNR-INFM), CASTI Regional Laboratory, 67100 L'Aquila, Italy\\
$^4$Physikalisches Institut, Universit\"at zu K\"oln, Z\"ulpicher Str. 77, 50937 K\"oln, Germany\\
$^5$Institute for Molecules and Materials, Radboud Universiteit, 6500 GL Nijmegen, The Netherlands\\
$^6$Stanford Institute for Materials and Energy Sciences, Stanford University and SLAC, Menlo Park}

\begin{abstract} 
We show that charge ordered rare-earth nickelates of the type RNiO$_3$ (R= Ho, Lu, Pr and Nd) are multiferroic with very large magnetically induced ferroelectric (FE) polarizations. This we determine from first principles electronic structure calculations.  The emerging FE polarization is directly tied to the long-standing puzzle of which kind of magnetic ordering is present in this class of materials: its direction and size indicate the type of ground-state spin configuration that is realized. Vice versa, the small energy differences between the different magnetic orderings suggest that a chosen magnetic ordering can be stabilized by cooling the system in presence of an electric field. 
\end{abstract}

\date{\today} 


\maketitle

{\it Introduction}
Complex oxides with simultaneous magnetic and ferroelectric (FE) ordering --multiferroics-- are attracting enormous scientific interest~\cite{Cheong07,Khomskii09}. They offer the potential to control the magnetic order parameter by the FE one and vice versa --a very desirable property from a technological point of view~\cite{Eerenstein06}. Even if in quite a number of transition metal oxides both ferroelectricity and magnetism are present, magnetically induced FE polarizations observed so far are typically very small~\cite{Kimura03,Hur04,Giovannetti08,Lottermoser04}. This smallness is particularly pronounced in materials where multiferroicity relies on relativistic spin-orbit coupling, which is intrinsically weak~\cite{Cheong07}. 

A few years ago it was pointed out that, theoretically at least, materials that are simultaneously {\it magnetic} and {\it charge ordered} can be multiferroic and potentially have a very large polarization~\cite{Efremov04,Brink08}. To become multiferroic, however, an insulting oxide needs to meet an additional requirement: its symmetry has to be such that magnetic ordering can push a charge ordering pattern from {\it site-centered} towards {\it bond-centered}~\cite{Efremov04}. A large polarization results if the oxide is in addition electronically soft, so that inside it charge can easily be displaced.

Here we show that precisely this scenario materializes in perovskite nickelates RNiO$_3$, where R is a rare earth element such as Ho, Lu, Pr or Nd. Consequently these nickelates can exhibit magnetically induced FE polarizations ($\bf{P}$) that are very large, up to 10 $\mu C/cm^2$. Such a polarization is two orders of magnitude larger than the one of  typical multiferroics such as TbMnO$_3$\cite{Kimura03} or TbMn$_2$O$_5$~\cite{Hur04}. Also a very interesting fundamental point is associated with the symmetry of $\bf{P}$ in the rare-earth nickelates. To appreciate this aspect we have to bear in mind that in spite of their apparently simple chemical formula, the rare-earth nickelates are very complex materials. They show an intriguing and only partially understood transition from a high temperature metallic phase into a low temperature insulating one. The nature of magnetic order in this low temperature insulating phase has been a long standing puzzle. Three different magnetic structures have been proposed, two of which are collinear and one non-collinear, and so far experiments have not been able to differentiate between them.

We show that all the proposed magnetic structures of the rare-earth nickelates are similar in the sense that all are multiferroic and very close in energy. However, different magnetic symmetries leave an individual fingerprint on the {\it size} and, in particular the {\it direction} of ferroelectric polarization $\bf{P}$. In one type of collinear magnetic ordering, for instance, $\bf{P}$ is parallel to the crystallographic $b$-axis; in the other it is perpendicular to this axis. With this theoretical result in hand, an experimental determination of the direction and magnitude of $\bf{P}$ will reveal the precise type of magnetic ordering that is realized in the rare-earth nickelates. This fundamental observation also suggests a practical application:  in these nickelates it allows to control the realization of different magnetic phases by cooling the material through its magnetic phase transition in an externally applied electric field.

\begin{figure}
\centerline{\includegraphics[width=\columnwidth,angle=0]{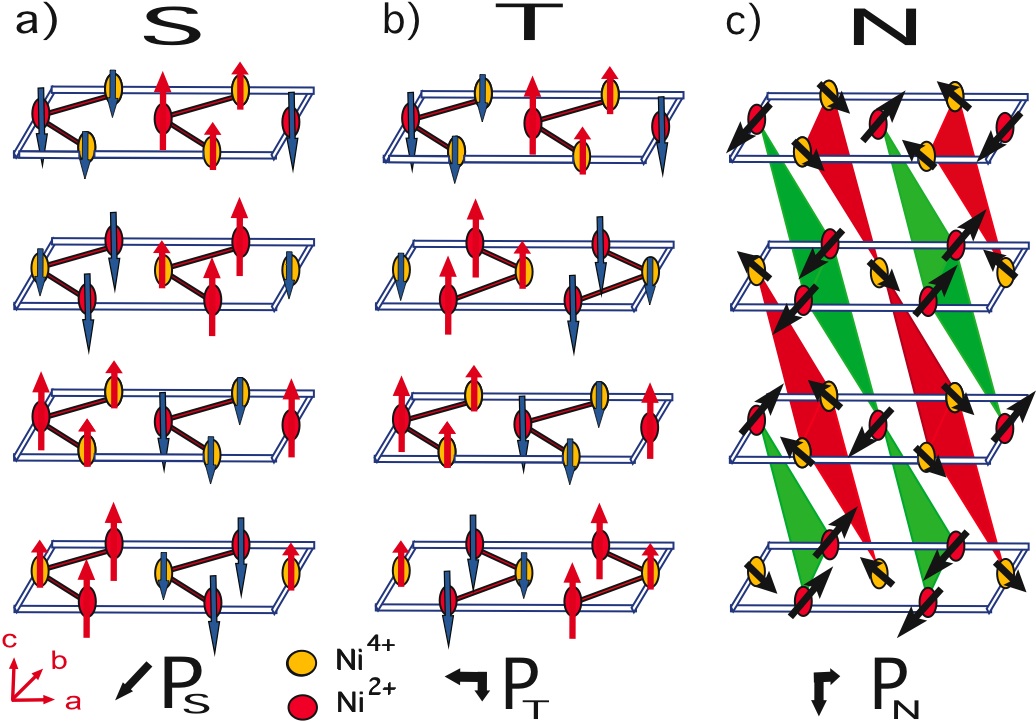}}
\caption{(Color online) Schematic view of the charge and magnetic ordered a) {\it S-type}  b) {\it T-type} and c) {\it N-type} non-collinear magnetic structure of RNiO$_3$.  Ni$^{2+}$ and Ni$^{4+}$ are shown in red and orange. Arrows represent Ni spins. In a) and b) the zig-zag spin chains are indicated. In c) the FM planes perpendicular to [101] are highlighted. The vector $\bf P$ shows the direction of polarization obtained in our calculations.}
\label{fig1}
\end{figure}

{\it Lattice and Charge Order} The metal-insulator transition in RNiO$_3$ (R=Pr, Nd, Sm, Ho, Lu etc.) takes place at relatively high temperatures: $T_{MI}=$130 K (Pr), 200 K (Nd), 400 K (Sm), 580 K (Ho) and 600 K (Lu) and is believed to coincide with the appearance of charge ordering and a simultaneous transition of the crystallographic symmetry from orthorhomic Pbnm to monoclinic P2$_1$/n~\cite{Alonso00,Medarde08}.  The charge ordering is characterized by a nickel charge disproportionation of formally Ni$^{3+}$ into Ni$^{(3+\delta)+}$ and Ni$^{(3-\delta)+}$, which form a simple two-sublattice, rocksalt-like, superstructure, see Fig.~\ref{fig1}. To be specific we will consider in the following four representative members from the nickelate series: R = Ho, Lu, Pr and Nd.  Of these R=Ho and Lu are small rare-earth ions, with a magnetic ordering temperature $T_{\rm N}$ below the metal-insulator transition: $T_{\rm N} < T_{\rm MI}$ (for Ho $T_{\rm N}$=145 K, for Lu 130 K).  The nickelates with the larger  rare-earth ions, Pr or Nd, are different in that their metal-insulator transition coincides with the appearance of magnetic ordering: T$_{\rm N}=$T$_{\rm MI}$.  The monoclinic P2$_1$/n  crystal structure contains two inequivalent Ni positions and three inequivalent oxygen atoms (O$_1$, O$_2$, O$_3$)~\cite{Alonso99,Alonso00,Medarde08}. Experimentally the charge ordering reflects itself in an oxygen breathing distortion of the NiO$_6$ octahedra and induces different magnetic moments on the two inequivalent Ni atoms (1.4/1.4 $\mu_B$ and 0.6/0.7 $\mu_B$ for Ho/Lu, respectively)~\cite{Diaz01,Mazin07}. 

{\it Magnetic Order} 
The magnetic neutron diffraction peaks are characterized by the propagation vector ${\bf Q}=({1 \over 2},0,{1 \over 2})$~\cite{Diaz01,Munoz94,Alonso99}. In this class of nickelates the experiments suggested three possible magnetic structures. In the following we label  two collinear magnetic structures as S, T , and the non-collinear one as N (see Fig.~\ref{fig1}).  In both the S and T-type magnetic structure the Ni spins within the $ab$-plane form ferromagnetic zig-zag chains. Adjacent zig-zag chains in the same plane are coupled antiferromagnetically. The stacking of the zig-zag chains along the $c$-axis differentiates between S and T-type ordering: in S-type (Fig.~\ref{fig1}a) the zig-zag spin-chains are all pointing in the same direction, whereas for T-type  (Fig.~\ref{fig1}b) the zig-zag chains in adjacent planes are pointing in alternate +$a$ and -$a$ directions. This implies for T-type that the spins in planes perpendicular to the [-1,0,1] direction are pointing in the same direction, see Fig.~\ref{fig1}b. Along this [-1,0,1] direction these ferromagnetic planes are ordered in a $\uparrow \uparrow \downarrow \downarrow \uparrow \uparrow \downarrow \downarrow$ fashion~\cite{Brink08}. In the non-collinear N-structure (Fig.~\ref{fig1}c~\cite{Scagnoli06}) all spins lie in the $ac$ plane, with spins in a plane perpendicular to [1,0,1] pointing in the same direction. When moving from plane to plane along [1,0,1] the spins rotate within the $ac$-plane. The N-type magnetic structure therefore corresponds to a spin-spiral.
\begin{table}[b]
\begin{center}
\begin{tabular}{|c ||c c c | c|| c c c | c  | }
\hline 
& \multicolumn{4}{c||}{experimental structure} &  \multicolumn{4}{c|}{relaxed structure}  \\  \hline
Rare &  \multicolumn{3}{c|}{T-type}&  \multicolumn{1}{c||}{S-type}
	  &  \multicolumn{3}{c|}{T-type}&  \multicolumn{1}{c|}{S-type}\\   
earth & $ P_{tot}$ & $P_a$  & $P_c$ &  $P_b$  
         & $ P_{tot}$ & $P_a$  & $P_c$ &  $P_b$ \\
\hline
Lu 	 & 10.31& 9.91& 2.84 & 5.21 & 9.86& 9.82& 0.76 & 7.07\\
Ho 	 &  8.66& 8.05& 3.19 &3.60  & 10.46&10.38& 1.39 & 6.91\\
Pr 	 & 14.80& 13.23& 6.64 & 1.81 & 7.87& 7.82& 0.94 & 2.57\\ \hline
Nd 	 & & & &  & 8.38& 8.28& 1.29 & 3.13 \\
\hline
\end{tabular}
\end{center}
\caption{FE polarization ${\bf P}$ of RNiO$_3$ (R=Ho, Lu, Pr) in $\mu C/cm^2$ for both the experimental centrosymmetric (Ref.~\cite{Alonso00} for Ho, Lu and Ref.~\cite{Medarde08} for Pr) and the relaxed crystal structure, for either S or T-type magnetic ordering, with P$2_1$ and P$n$ symmetry, respectively. Atomic positions for NdNiO$_3$ taken from experiments in the $Pbnm$ orthorhombic setting~\cite{Munoz92} and then relaxed without changing the Bravais lattice.}
\label{table1}
\end{table}

{\it Ab Initio Results}  
For the crystal and magnetic structures outlined above we performed a set of density functional calculations using the projector augmented-wave (PAW) method and  a plane-wave basis sets as implemented in VASP \cite{VASP}. We include the strong Coulomb interactions between the Ni $3d$ electrons,  in SGGA+U \cite{PW91,PAW+U,Liechtenstein95} calculations for $U$=8 eV and a Hund's rule exchange of $J_H=0.88$ eV. Starting from the experimental centrosymmetric crystal structures we compute the electronic structure for T, S and N magnetic order.  In all the calculations, the difference in total energy between the T and S type ordering is very small, within the numerical accuracy.  We evaluate the electronic contributions to the polarization with the Berry phase method within the PAW formalism \cite{BerryPhase,notaintegration}. In Table \ref{table1} we report the electronic contributions to the FE polarization for the T and S type magnetic ordering. For the non-collinear N-type ordering these calculations are numerically extremely demanding. We have therefore computed the polarization for HoNiO$_3$ in its experimental crystal structure only, resulting in  $P_c$=-110 $nC/cm^2$ and $P_a$=20 $nC/cm^2$. It is remarkable that for T-type the polarization is large, in the $ac$-plane and predominantly along the $a$-axis, for S-type it is large and strictly along the $b$-axis and for N-type it is weak, in the $ac$-plane and mostly along $c$.

\begin{figure}
\centerline{\includegraphics[height=.53\columnwidth,angle=0]{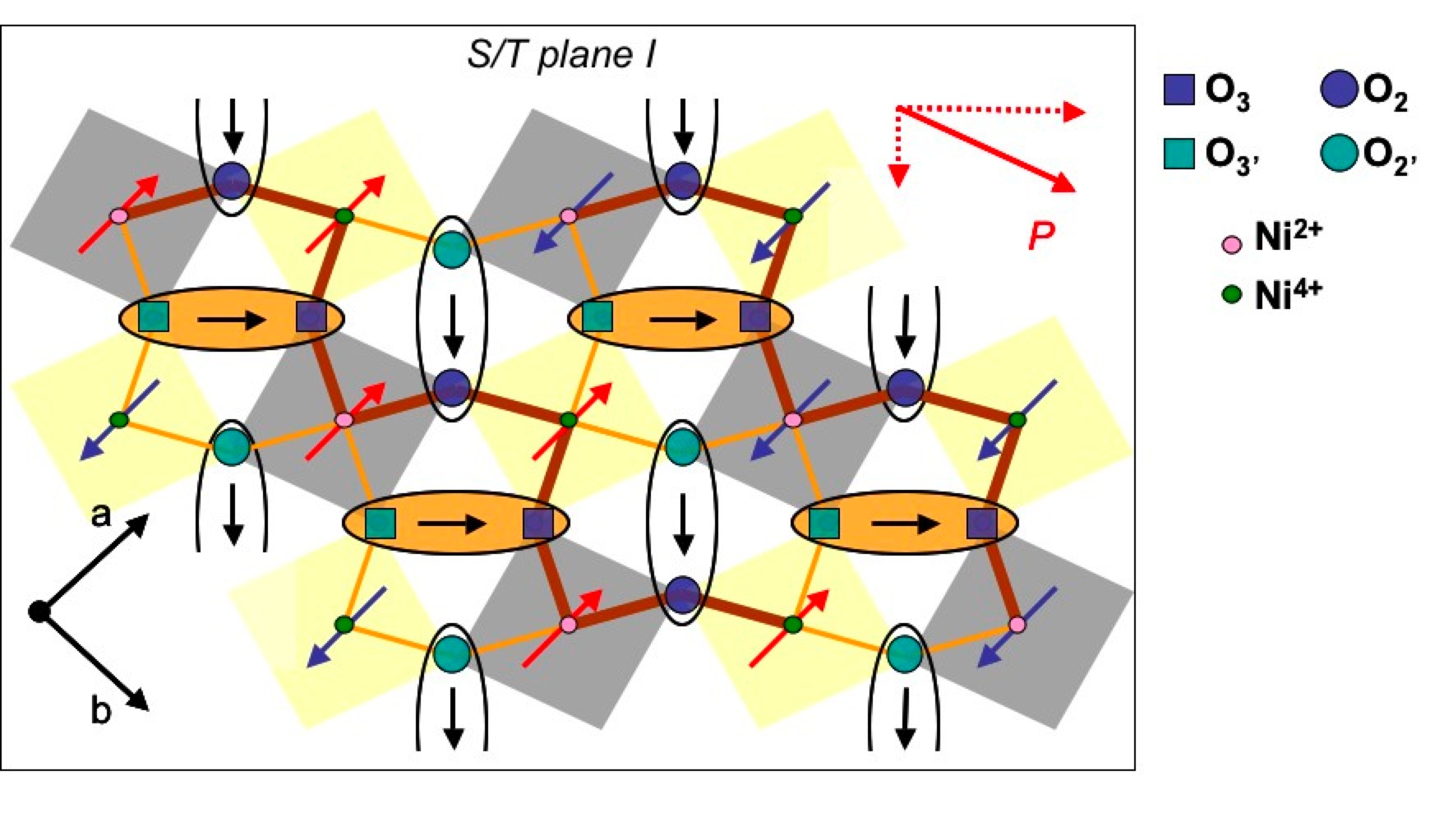}}
\caption{(Color online) Schematic arrangement of oxygen-induced dipoles and resulting FE polarization for the P$2_1$/n crystal structure, in the top layer of Fig.~\ref{fig1}a and b.
}
\label{fig2}
\end{figure}

{\it S- and T-type Magnetism: Origin of Multiferroicity}
Having established that the magnetic ordering induces a very significant FE polarization in the rare-earth nickelates, we now clarify its microscopic origin and explain its different direction for T and S-type order. In the P2$_1$/n structure the corner-sharing NiO$_6$ octahedra are distorted and tilted due to the so-called GdFeO$_3$ distortion, which causes a counter-rotation of neighboring octahedra in the $ab$-plane, see Fig.~\ref{fig2}. Neighboring planes along the $c$-axis display a similar in-plane oxygen displacement pattern. The end result of the distortions is that each Ni ion is surrounded by two crystallographically inequivalent oxygen sites in the $ab$-plane: O$_2$ and O$_3$, see Fig.~\ref{fig2}. Along one direction chains of O$_2$-Ni-O$_2$-Ni bonds form and perpendicular to it, all bonds are O$_3$-Ni-O$_3$-Ni. In spite of these distortions the crystal structure is still centrosymmetric and the material therefore paraelectric. However, the formation of zig-zag {\it spin} chains in the $ab$-plane breaks this inversion symmetry. 

The magnetic ordering along the Ni-O$_2$ chain direction can be denoted as Ni$_\uparrow$-O$_2$-Ni$_\uparrow$-O$_2$-Ni$_\downarrow$. In this structure the O$_2$ sites become inequivalent as one O$_2$ is in between charge disproportionate Ni ions with parallel spin and the other one between antiparallel Ni spins. Thus the oxygen sites split into O$_2$ and O$_{2^\prime}$, see Fig.~\ref{fig2}. Besides the charge disproportionation on the Ni sites, now also the inequivalent oxygen atoms  O$_2$ and O$_{2^\prime}$ charge-polarize. For the Ni-O$_3$ chain the situation is similar and the splitting is into O$_3$ and O$_{3^\prime}$. The inequivalence of the four in-plane oxygens ions surrounding a Ni is directly reflected by their different Born effective charges $Z^*$: when for instance R=Lu we find $Z^*(O_{2^\prime})=-1.25e$, $Z^*(O_2)=-6.37e$, $Z^*(O_{3^\prime})=-1.59e$ and $Z^*(O_3)=-1.73e$. For other rare-earths we observe similar trends. The resulting inequivalence of oxygen ions  situated on the nickel bonds causes a partial shift away from a nickel site-centered charge ordering to a Ni-Ni bond centered charge ordering. The resulting net dipole moment of each nickel-oxide $ab$-plane has a finite projection along both the $a$ and $b$-axis, see Fig.~\ref{fig2}. This dipole formation is reminiscent of the mechanism for ferroelectricity proposed in Ref.~\cite{Picozzi07}, where it was discussed in the context of HoMnO$_3$. 
\begin{figure}
\includegraphics[height=.5\columnwidth,angle=0]{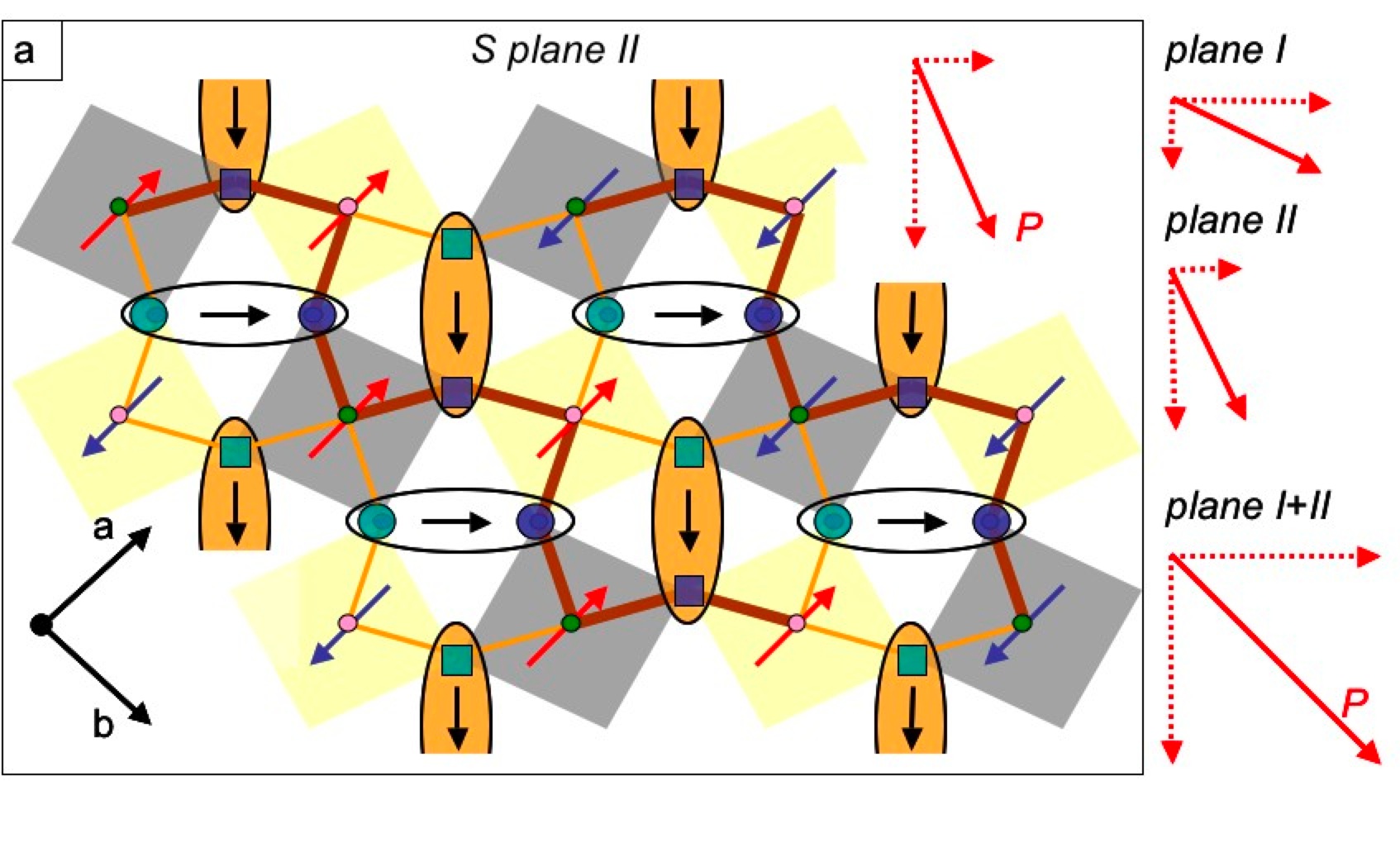}
\includegraphics[height=.53\columnwidth,angle=0]{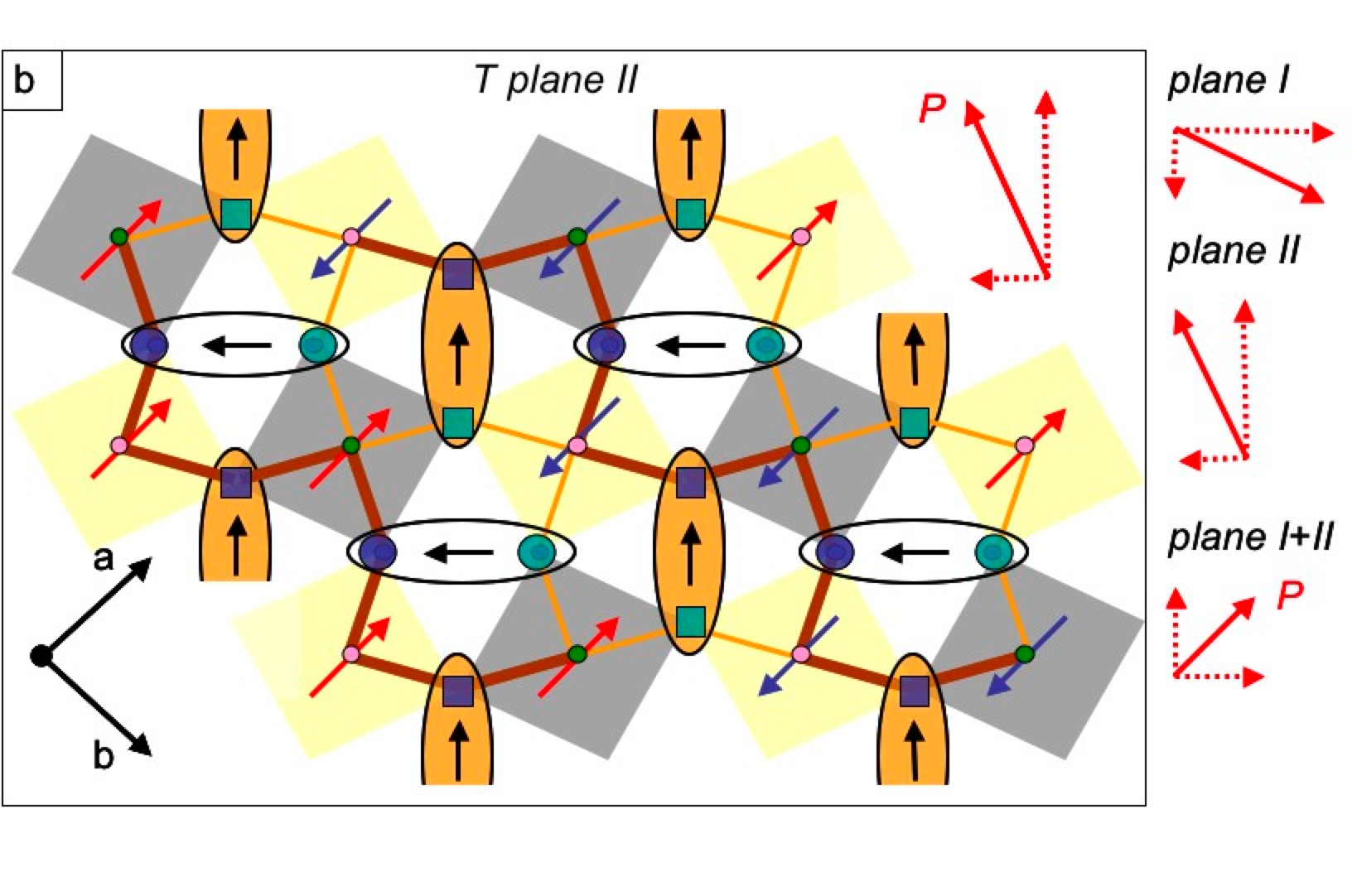}
\caption{(Color online) Ferroelectric polarization after an elementary translation along the $c$-axis, corresponding to the second layer Fig.~\ref{fig1}, for a)  S- and b) T-type magnetic ordering.
}
\label{fig3}
\end{figure}
For S-type magnetic order the dipole moments of different nickel-oxide $ab$-planes add up to a net polarization along the $b$-axis. The following symmetry argument underlies the fact the polarization vanishes in all other directions. Translating an $ab$ plane along [001] interchanges the $O_2$ and $O_3$ positions, but leaves alternation pattern of parallel/anti-parallel spin bonds invariant, see Fig.~\ref{fig3}a. The $O_2$-$O_3$ interchange is equivalent to a rotation around the $b$-axis  by $\pi$. As in this case also $\bf P$ rotates by $\pi$ around $b$-axis, the net polarization points fully along $b$, see Fig.~\ref{fig3}a. 

By virtue of the same argument for T-type magnetic ordering the polarization along the $b$-axis vanishes, see Fig.~\ref{fig3}b. In this case a translation along [001] again interchanges the $O_2$ and $O_3$ positions, but the bonds of parallel spins transform into antiparallel ones and vice versa. This implies that compared to the S-type structure, an elementary translation along [001] in T-type generates an additional reflection of $\bf P$ in the $a$-axis, see Fig.~\ref{fig3}b. Consequently the net in-plane polarization of the system is now along $a$. Via a similar argument one finds in T-type a finite polarization along the $c$-axis, which is forbidden in the S-type structure.

{\it  Non-collinear phase: Origin of Multiferroicity} 
The N-type structure corresponds to a magnetic spiral with propagation vector $\bf Q$= [1/2,0,1/2] and rotation axis $\bf e$ = [0,1,0]: it is a spiral with spins rotating in the $ac$-plane. The spiral structure implies that in the direction of $\bf Q$ the {\it angle} between spins in successive planes is invariant, so that magnetic bonds are equivalent. The N-type magnetic ordering can therefore not give rise to multiferroicity via a modulation of the charge ordering, as is the case in the S- and T-phase. However, in spin-spirals the relativistic spin-orbit coupling directly causes a FE polarization~\cite{Cheong07}. Symmetry dictates that this polarization arises in the direction ${\bf e} \times {\bf Q}$, corresponding to [1,0,-1] in our case. Indeed,  our {\em ab-initio} calculations including spin-orbit coupling give ${\bf P} =(20, 0, -110) \ nC/cm^2$, in full agreement with these symmetry considerations. The size of the polarization is comparable to that of TbMnO$_3$~\cite{Kimura03} and thus much weaker than that for the S- and T-type ordering, which is a generic feature of systems where multiferroicity is caused by spin-orbit coupling.

{\it Relaxed structure}
We have also computed the ionic contribution to $\bf P$ that arises from lattice distortions induced in the FE phase. Starting from the experimental centrosymmetric P2$_1$/n crystal structure with a magnetic supercell we relax unit cell ionic positions, allowing for a lower symmetry structure to develop. Details of the structural relaxations will be published elsewhere~\cite{long_paper}. The resulting total FE polarizations are reported in Table~\ref{table1}. For the S-type magnetic state the induced lattice distortions enhance $\bf P$ for all nickelates that we have considered. For T-type also a reductions of the polarization occurs in some materials, but still $\bf P$ stays very large: $\sim10  \mu C/cm^2$ for Ho/Lu,   $\sim8 \mu C/cm^2$ for Pr/Nd. For T-type the polarization along the $a$-axis, $P_a$,  dominates over $P_c$. We find that $P_c$ is so small due to a partial cancelation of ionic and electronic contributions to $P_c$, an effect that was also observed in HoMn$_2$O$_5$~\cite{Giovannetti08} and in TbMnO$_3$~\cite{Malashevich08}. For the S-type structure we find total $P_b$ $\sim7  \mu C/cm^2$ for Ho/Lu,   and $\sim3 \mu C/cm^2$ for Pr/Nd. The generic observation is that nickelates with small rare-earth ions (Ho and Lu) tend to show the largest polarizations. 

{\it Conclusions} On the basis of theoretical calculations we predict that the perovskite nickelates RNiO$_3$ (R - rare earth) are multiferroic in their low-temperature insulating magnetic phase. We show there are different mechanisms at play for magnetically-induced ferroelectricity  in RNiO$_3$. The S and T collinear magnetic spin configurations give a remarkably large polarization ($\sim10$ $\mu$C/cm$^2$) along the $b$ axis and in the $ac$ plane, respectively. It is driven by non-centrosymmetric arrangements of spins which force the charge-ordering  to shift from site-centered to partially bond-centered.  The estimated polarization in the S and T collinear cases is much larger than that arising from a relativistic spin-orbit related mechanism, which is at play in the N-type spin-spiral state.  An experimental determination of the direction and magnitude of $\bf P$ can therefore solve the long-standing puzzle related to the magnetic ground-state in nickelates.   The fact that according to our calculations the energies of the S- and T-type magnetic structures are very close  but the directions of polarization are quite different, suggests one can stabilize one or the other by cooling in  an appropriate electric field, so that in effect an applied external electric-field can control the realization of different magnetic phases in these nickelates.

{\it Acknowledgements} We thank Maxim Mostovoy for stimulating discussions.  This work is supported by Stichting FOM, NCF and NanoNed, The Netherlands and by SFB 608, Germany. The research leading to part of these results has received funding from the European Research Council under the European Community Seventh Framework Program (FP7/2007-2013)/ERC Grant Agreement No. 203523-BISMUTH.


\begin{thebibliography}{99}
\bibitem{notaintegration} The cut-off for the plane-wave basis set was 400 eV and a [4,8,2]  mesh was used for the  Brioullin-zone sampling. In the BP approach, we integrated the center of the Wannier functions by strings parallel to all three axes, each of them divided into 8 k-points.
\end{thebibliography}
\end{document}